\documentclass[12pt,thmsa]{article}
\usepackage{amssymb}

\usepackage{sw20aip}

\input{tcilatex}
\begin{document}

\author{S.B. Rutkevich\thanks{%
Institute of Solid State and Semiconductor Physics, P. Brovka str., 17,
Minsk 220072, Belarus, e-mail: rut@ifttp.bas-net.by}}
\title{Analytic Verification of the Droplet Picture in the Two-Dimensional Ising
Model}
\date{}
\maketitle

\begin{abstract}
It is widely accepted that the free energy $F(H)$ of the two-dimensional
Ising model in the ferromagnetic phase $T$$<$$\,T_{c}$ has an essential
branch cut singularity at the origin $H=0$. The phenomenological droplet
theory predicts that near the cut drawn along the negative real axis $H$$<$$0
$, the imaginary part of the free energy per lattice site has the form $%
\mathrm{Im}F[\,\exp (\pm i\pi )\mid H\mid \,]=\pm B|H|\exp (-A/\mid H\mid )$
for small $\mid H\mid $ . We verify this prediction in analytical
perturbative transfer matrix calculations for the square lattice Ising model
for all temperatures $0$$<$$T$$<$$T_{c}$ and arbitrary anisotropy ratio $%
J_{1}/J_{2}$. We obtain an expression for the constant $A$ which coincides
exactly with the prediction of the droplet theory. For the amplitude $B$ we
obtain $B=\pi M/18$, where $M$ is the equilibrium spontaneous magnetization.
In addition we find discrete-lattice corrections to the above mentioned
phenomenological formula for $\mathrm{Im}F$, which oscillate in $H^{-1}$.%
\medskip \newline
\newline
\end{abstract}

\section{Introduction}

It is well known$^{(\citenum{Is}-\citenum{AU})}$, that in the ferromagnetic
phase $0<T$$<$$\,T_{c}$ the free energy of the two-dimensional Ising model
as the function of the magnetic field $H$ has a so-called droplet
singularity at the origin $H=0$. This singularity prevents analytical
continuation of the free energy from positive to negative values of $H$
along the real $H$-axis. The phenomenological droplet (nucleation) theory$^{(%
\citenum{A}-\citenum{Ka})}$ claims, however, that the free energy can be
continued from positive to negative magnetic fields along a circle going
around the origin in the complex $H$-plane (see Fig. 1). According to this
theory, the free energy per lattice site $F(H)$ continued in such a way
gains on the negative real axis $H<0$ a nonzero imaginary part, which is
expected to have the form 
\begin{equation}
\mathrm{Im}F[\,\exp (\pm i\pi )\mid H\mid \,]=\pm B\,|H|\exp (-A/\mid H\mid )
\label{FF}
\end{equation}
for small $\mid H\mid $ . The sign of this imaginary part depends on the
side, from which one approaches to the negative real axis $H<0$. Expression (%
\ref{FF}) extrapolates to the ferromagnetic Ising model the results obtained
in the semiclassical nucleation field theory analysis of the coarse-grained
Ginzburg-Landau model.$^{(\citenum{Langer}-\citenum{CC})}$ In the nucleation
theory, the free energy continued to the cut $H<0$ is interpreted as the
free energy of the metastable state: 
\[
F_{ms}(H)\equiv F(e^{i\pi }\mid H\mid ). 
\]
Langer conjectured,$^{(\citenum{Langer})}$ that $\mathrm{Im}F_{ms}(H)$ may
be identified (up to a dynamical factor) with the metastable phase decay
rate provided by the thermally activated nucleation of the critical droplet.

The phenomenological droplet theory prediction for the amplitude $A$ in (\ref
{FF}) is$^{(\citenum{RG})}$ 
\begin{equation}
A=\frac{\beta \hat{\Sigma}^{2}}{8M},  \label{fd}
\end{equation}
where $M$ is the spontaneous magnetization, and $\hat{\Sigma}^{2}\ $denotes
the square of surface free energy of the equilibrium-shaped droplet divided
by its area. Both $\hat{\Sigma}^{2}$ and $M$ relate to the equilibrium
zero-field state, and are known exactly. The linear depending on $\left|
H\right| $ prefactor in (\ref{FF}) arises in the continuum droplet field
theory$^{(\citenum{Langer}-\citenum{ZW})}$ from the contribution of the
surface excitations of the critical droplet. Voloshin claimed$^{(%
\citenum{Voloshin2})}$ that, if fluctuations are continuum and isotropic,
the prefactor in (\ref{FF}) becomes universal. Extrapolation of the
Voloshin's continuum droplet field theory result to the $d=2$ Ising model
leads to the following prediction for the amplitude $B:$%
\begin{equation}
B\mapsto B_{V}=\frac{M}{2\pi }.  \label{Vol}
\end{equation}

In the continuum field theory, the imaginary part of the free energy appears
in the functional integral calculations. In the alternative approach to the
droplet singularity problem, one deals with eigenvalues of the Ising model
transfer-matrix. Numerical transfer-matrix calculations initiated by Privman
and Schulman$^{(\citenum{Pr})}$ and continued by G\"{u}nther, Rikvold and
Novotny$^{(\citenum{GRN1},\citenum{GRN})}$ confirm equations (\ref{FF}) and (%
\ref{fd}). These equations were confirmed also by Lowe and Wallace,$^{(%
\citenum{LW})}$ and by Harris$^{(\citenum{Ha})}$ in numerical analysis of
the small-$H$ power expansion for the magnetization $M(H).$ Recently
analytic transfer-matrix derivation of equations (\ref{FF}), (\ref{fd}) for
the $d=2$ Ising model has been done$^{(\citenum{R})}$ in the extreme
anisotropic limit.

In this paper we generalize the transfer matrix approach developed in paper$%
^{(\citenum{R})}$ and verify analytically the droplet theory predictions (%
\ref{FF}), (\ref{fd}) for the square lattice Ising model for all
temperatures $0$$<$$T$$<$$T_{c}$ and arbitrary anisotropy ratio $J_{1}/J_{2}$%
. We obtain an expression for the constant $A$ which coincides exactly with
the prediction of the droplet theory. For the amplitude $B$ we find $B=\pi
M/18$, which is very close to Voloshin's result (\ref{Vol}): $B/B_{V}=\pi
^{2}/9\approx 1.0966$. We suppose, that this small discrepancy results from
approximations used in our calculations. Obtained values for the amplitude $%
B $ are compared with those extracted numerically from the known
coefficients of the expansion of the magnetization in powers of $H$ by means
of dispersion relations.$^{(\citenum{LW,1})}$

We find also the discrete-lattice corrections to the phenomenological
formula (\ref{FF}), which oscillate in $H^{-1}$. The period of oscillations
agrees well with that observed by G\"{u}nther \textit{et al.}$^{(%
\citenum{GRN})}$ in numerical constrained transfer matrix calculations.

\section{Transfer matrix and Hamiltonian\label{tr}}

The nearest neighbor Ising model on the square lattice in the magnetic field 
$H$ is defined by the energy 
\begin{equation}
\mathcal{E}=-\sum_{n=1}^{\mathcal{N}}\sum_{m=1}^{\mathcal{M}}\left(
J_{1}\sigma _{m,n}\sigma _{m+1,n}+J_{2}\sigma _{m,n}\sigma _{m,n+1}+H\sigma
_{m,n}\right)  \label{En}
\end{equation}
where $\sigma _{m,n}=\pm 1,$ the first/second index of $\sigma _{m,n}$
specifies the row/column of the lattice, $\mathcal{M}$ and $\mathcal{N}$
denote the number of rows and columns in the lattice, respectively. Periodic
boundary conditions are implied.

The row to row transfer matrix may be defined as $\hat{T}=e^{U}\hat{T}_{2}\ 
\hat{T}_{1},$ where 
\begin{eqnarray}
\hat{T}_{1} &=&\left[ 2\sinh (2K\,_{1})\right] ^{\mathcal{N}/2}\exp \left(
K\,_{1}^{*}\sum_{n=1}^{\mathcal{N}}\sigma _{n}^{1}\right) ,\quad \hat{T}%
_{2}=\exp \left( K\,_{2}\sum_{n=1}^{\mathcal{N}}\sigma _{n}^{3}\sigma
_{n+1}^{3}\right) ,\quad  \nonumber \\
U &=&h\sum_{n=1}^{\mathcal{N}}\,\sigma _{n}^{3}.  \label{U}
\end{eqnarray}
Here we have used the standard notations 
\[
K_{1}=\beta J_{1},\quad K_{2}=\beta J_{2},\quad h=\beta H,\quad
2K\,_{1}^{*}=-\ln \left( \tanh K_{1}\right) , 
\]
$\beta $ is the inverse temperature, $\sigma _{n}^{\alpha }\ (\alpha =1,2,3)$
are the Pauli matrices relating to the cite $n$ in the row.

The transfer matrix may be chosen in the symmetric form 
\[
\hat{T}_{S}=\left[ \hat{T}_{S}^{(0)}\right] ^{1/2}e^{U}\ \left[ \hat{T}%
_{S}^{(0)}\right] ^{1/2}, 
\]
where $\hat{T}_{S}^{(0)}$ is the symmetric transfer matrix of the Ising
model in zero magnetic field: 
\[
\hat{T}_{S}^{(0)}=\hat{T}_{2}^{1/2}\hat{T}_{1}\hat{T}_{2}^{1/2}. 
\]
As it was shown by Schultz, Mattis and Lieb,$^{(\citenum{SML})}$ the latter
becomes diagonal in fermionic variables: 
\begin{eqnarray}
\hat{T}_{S}^{(0)} &=&C\exp \left( -\mathcal{H}^{(0)}\right) ,\qquad 
\nonumber \\
\mathcal{H}^{(0)} &=&\int_{-\pi }^{\pi }\frac{\text{d}\theta }{2\pi }\
\omega (\theta )\ \psi ^{\dagger }(\theta )\ \psi (\theta ),  \label{H0}
\end{eqnarray}
where $\mathcal{H}^{(0)}$ is the zero-field Hamiltonian, $\theta $ is the
quasimomentum, $C$ is an insufficient numerical factor. Fermionic operators $%
\psi ^{\dagger }(\theta )$, $\ \psi (\theta )$ satisfying the canonical
anticommutational relations 
\[
\left\{ \psi (\theta )\ ,\psi (\theta ^{\prime })\right\} =\left\{ \psi
^{\dagger }(\theta )\ ,\psi ^{\dagger }(\theta ^{\prime })\right\} =0,\qquad
\left\{ \psi ^{\dagger }(\theta )\ ,\psi (\theta ^{\prime })\right\} =2\pi
\delta (\theta -\theta ^{\prime }) 
\]
can be expressed in terms of the initial Pauli matrices by use of the
Jordan-Wigner and duality transformation (see Appendix \ref{apa}). The
fermionic spectrum $\omega (\theta )$ is given by 
\begin{eqnarray}
\exp \omega (\theta ) &=&\cosh 2K\,_{1}^{*}\cosh 2K\,_{2}-\cos \theta \sinh
2K\,_{1}^{*}\sinh 2K\,_{2}+  \label{eps} \\
&&\left[ \left( \cosh 2K\,_{1}^{*}\cosh 2K\,_{2}-\cos \theta \sinh
2K\,_{1}^{*}\sinh 2K\,_{2}\right) ^{2}-1\right] ^{1/2}.  \nonumber
\end{eqnarray}
Operator $U$ defined by (\ref{U}) also can be represented in the
thermodynamic limit $\mathcal{N\rightarrow \infty }$ in the fermionic
variables: 
\begin{equation}
U=hM\sum_{n\in \Bbb{Z}}:\exp \frac{\rho _{n}}{2}:,  \label{v}
\end{equation}
where 
\begin{eqnarray}
\frac{\rho _{n}}{2} &=&-\sum_{j<\,n}\psi _{j}^{(+)}\ \psi _{j}^{(-)},
\label{111} \\
\psi _{j}^{(+)} &=&i\int_{-\pi }^{\pi }\frac{d\theta }{2\pi }\ \frac{\exp
(ij\theta )}{\epsilon (\theta )}\left[ \psi (\theta )+\psi ^{\dagger
}(-\theta )\right] ,  \label{112} \\
\psi _{j}^{(-)} &=&i\int_{-\pi }^{\pi }\frac{d\theta }{2\pi }\ \exp
(ij\theta )\ \epsilon (\theta )\left[ -\psi (\theta )+\psi ^{\dagger
}(-\theta )\right] ,  \label{113} \\
\epsilon (\theta ) &=&\left( \frac{z_{1}+z_{1}^{-1}-2\cos \theta }{%
z_{2}+z_{2}^{-1}-2\cos \theta }\right) ^{1/4},  \nonumber \\
z_{1} &=&\tanh K\,_{1}^{*}/\tanh K\,_{2},\quad z_{2}=\tanh K\,_{1}^{*}\
\tanh K\,_{2},  \label{zz}
\end{eqnarray}
and $M$ is the zero-field magnetization. In the ferromagnetic phase $%
M=\left[ 1-k^{2}\right] ^{1/8},$ and $k<1,$ where $k=\left( \sinh
2K\,_{1}\sinh 2K\,_{2}\right) ^{-1}$ $.$ We have used in (\ref{v}) the
conventional notation $:...:$ for the normal ordering with respect to the
fermionic operators $\psi (\theta ),\ \psi ^{\dagger }(\theta )$. Derivation
of equations (\ref{v})-(\ref{113}) is described in Appendix \ref{apa}, in
the main points of which we follow Jimbo \textit{et al.}$^{(\citenum{JM})}$%
\textit{\ }

At zero magnetic field, the Hamiltonian $\mathcal{H}^{(0)}$ of the Ising
model is given by (\ref{H0}). Two ferromagnetic ground states $\mid
0_{+}\rangle $ and $\mid 0_{-}\rangle $ coexist in the ferromagnetic phase $%
k<1$. They are distinguished by the sign of the spontaneous magnetization $%
\langle 0_{\pm }\mid \sigma _{n}^{z}$ $\mid 0_{\pm }\rangle =\pm $ $M$. The
state $\mid 0_{+}\rangle $ characterized by the positive magnetization $+M$
is the ferromagnetic vacuum of $\psi (\theta )$-operators: $\psi (\theta
)\mid 0_{+}\rangle =0$ for all $\theta $.

A small magnetic field $h\neq 0$ changes the Hamiltonian $\mathcal{H}^{(0)}$
to 
\begin{equation}
\mathcal{H}(h)=-\ln \left( e^{-\mathcal{H}^{(0)}/2}e^{U}\ e^{-\mathcal{H}%
^{(0)}/2}\right) ,\quad \mathcal{H}(0)=\mathcal{H}^{(0)}.  \label{hami}
\end{equation}
It can be expanded in powers$\footnote{%
In the extreme anisotropic limit$^{(\citenum{R})}$ the Hamiltonian expansion
(\ref{h1}) containes only two terms: $\mathcal{H}(h)=$ $\mathcal{H}^{(0)}+%
\mathcal{H}^{(1)},$ where $\mathcal{H}^{(1)}=-U$.}$ of $h$: 
\begin{equation}
\mathcal{H}(h)=\sum_{j=0}^{\infty }\mathcal{H}^{(j)},  \label{h1}
\end{equation}
where $\mathcal{H}^{(j)}\sim h^{j}$.

\section{Modified perturbation theory}

Let us consider the eigenvalue problem 
\begin{equation}
\mathcal{H}(h)\ \mid \phi _{+}(h)\rangle =E(h)\mid \phi _{+}(h)\rangle ,
\label{eig}
\end{equation}
where $\mid \phi _{+}(0)\rangle =\mid 0_{+}\rangle $.

If $h>0$, the eigenvector $\mid \phi _{+}(h)\rangle $ is the ground state of
the Hamiltonian (\ref{hami}), and the corresponding energy $E(h)$ is
directly related with the Ising model free energy per lattice cite $%
F(h,\beta )$: 
\begin{equation}
F(h,\beta )=F(0,\beta )+\frac{E(h)}{\beta \mathcal{N}}.  \label{fr}
\end{equation}
The energy can be expanded in the formal power series 
\[
\quad E(h)=\sum_{j=1}^{\infty }h^{j}C_{j}, 
\]
which coefficients $C_{j}$ can be, in principal, determined from standard
Rayleigh-Schr\"{o}dinger perturbation theory.

However, if the magnetic field is small and negative $h<0$, the state $\mid
\phi _{+}(h)\rangle $ (with positive magnetization almost equal to $M$) must
be identified with the metastable (false) vacuum. It decays due to the
quantum tunneling, and the decay rate\footnote{%
Strictly speaking, the term ''decay rate'' here relates to the
quantum-mechanical model with Hamiltonian (\ref{h1}), but not to the initial
two-dimensional Ising model (\ref{En}). The nucleation rate in the latter
model contains also the so-called kinetic prefactor, which depends on the
detailed non-equilibrium dynamics.$^{(\citenum{RG})}$} $\Gamma $ is
proportional to the imaginary part of the energy $E(h)$ continued to
negative magnetic fields:$^{(\citenum{LL,RS})}$

\begin{equation}
\Gamma =-2\text{ Im }E(h).  \label{img}
\end{equation}
It turns out, however, that $\Gamma $ can not be determined from the
straightforward perturbation theory with the zero-order Hamiltonian $%
\mathcal{H}^{(0)}$. This is due to the fact that the term $\mathcal{H}^{(1)}$
in the expansion (\ref{h1}) contains the long-range interaction $(-U_{0})$
between fermions, which is given by 
\begin{equation}
-U_{0}\equiv -U\mid _{\epsilon (\theta )\rightarrow 1}=\mid h\mid
M\sum_{n\in \Bbb{Z}}:\exp \left( -2\sum_{j<\ n}b_{j}^{\dagger }b_{j}\right)
:\ .  \label{v0}
\end{equation}
Here 
\[
b_{j}^{\dagger }=\int_{-\pi }^{\pi }\frac{d\theta }{2\pi }\psi ^{\dagger
}(\theta )\exp \left( -ij\theta \right) ,\qquad b_{j}=\int_{-\pi }^{\pi }%
\frac{d\theta }{2\pi }\psi (\theta )\exp \left( ij\theta \right) , 
\]
are the operators which create/annihilate a fermion at the cite $j$.
Interaction (\ref{v0}) increases linearly with the distance between fermions,%
$^{(\citenum{R})}$ and therefore changes the structure of the Hamiltonian
spectrum. So, to describe decay of metastable vacuum $\ \mid \phi
_{+}(h)\rangle $, one should include the long-range interaction (\ref{v0})
into the zero-order Hamiltonian.

Accordingly, we subdivide the Hamiltonian (\ref{h1}) into the zero-order $%
\mathcal{H}_{0}$ and interaction $V$ parts, as follows: 
\begin{equation}
\mathcal{H}(h)=\mathcal{H}_{0}+V,  \label{h2}
\end{equation}
where 
\begin{eqnarray}
\mathcal{H}_{0} &\equiv &\mathcal{H}^{(0)}-U_{0}-\mathcal{N\,}\left|
h\right| \,M,  \label{0h} \\
\qquad V &\equiv &\mathcal{H}^{(1)}+U_{0}+\mathcal{N\,}\left| h\right|
\,M+\sum_{j=2}^{\infty }\mathcal{H}^{(j)}  \label{vt}
\end{eqnarray}
The numerical constant $\mathcal{N\,}\left| h\right| \,M$ in (\ref{0h}) is
chosen to provide $\mathcal{H}_{0}\mid 0_{+}\rangle =0$.

\subsection{Zero order spectrum}

Consider the zero-order eigenvalue problem 
\begin{equation}
\mathcal{H}_{0}\mid \phi _{l}\rangle =E_{l}\ \mid \phi _{l}\rangle
\label{zero}
\end{equation}
First note, that eigenstates $\mid \phi _{l}\rangle $ can be classified by
the fermion number, since the modified zero-order Hamiltonian (\ref{0h}){\ }%
conserves the number of fermions. As in paper,$^{(\citenum{R})}$ we shall
consider only two-fermions (i.e. one-domain) states in (\ref{zero}).
Physically, this means that we neglect interaction between nucleating
droplets of the stable phase.

In the coordinate representation equation (\ref{zero}) takes the form 
\[
\sum_{n^{\prime }\in \Bbb{Z}}K_{nn^{\prime }}\ \phi _{l}(n^{\prime })-M\mid
n\ h\mid \phi _{l}(n)=\frac{E_{l}}{2}\phi _{l}(n), 
\]
where 
\begin{eqnarray*}
K_{nn^{\prime }} &=&\int_{-\pi }^{\pi }\frac{d\theta }{2\pi }\ \omega
(\theta )\exp [i(n-n^{\prime })\theta ], \\
\phi _{l}(n) &=&\langle 0_{+}\mid b_{0}\ b_{n}\mid \phi _{l}\rangle ,\qquad
\phi _{l}(-n)=-\phi _{l}(n).
\end{eqnarray*}
If the energy $E_{l}$ is small enough $E_{l}\ll \omega (0)$, the
wavefunction $\phi _{l}(n)$ is mainly concentrated far from the origin in
the classically available region $\mid n\mid >\omega (0)/(\mid h\mid M)$.
Therefore, we can apply the `strong coupling approximation'$^{(\citenum{Zim}%
)}$ to represent the wavefunction in the form 
\begin{equation}
\phi _{l}(n)\simeq \varphi _{l}(n)-\varphi _{l}(-n),  \label{str}
\end{equation}
where the function $\varphi _{l}(n)$ solves the equation 
\[
\sum_{n^{\prime }\in \Bbb{Z}}K_{nn^{\prime }}\ \varphi _{l}(n^{\prime
})-\mid h\mid M\ n\ \varphi _{l}(n)=\frac{E_{l}}{2}\varphi _{l}(n). 
\]
\newline
\newline
After the Fourier transform, we obtain 
\[
\varphi _{l}(n)=\int_{-\pi }^{\pi }\frac{d\theta }{2\pi }\varphi _{l}(\theta
)\exp \left( in\theta \right) , 
\]
where 
\begin{eqnarray}
\varphi _{l}(\theta ) &=&C\ \exp \left\{ -\frac{i}{2\mid h\mid M}\left[
f(\theta )-E_{l}\ \theta \right] \right\} ,  \label{fii} \\
C &=&(2\mid h\mid M\ \mathcal{N})^{-1/2},  \nonumber \\
f(\theta ) &=&2\int_{0}^{\theta }\text{d}\chi \ \omega (\chi ).  \nonumber
\end{eqnarray}
The $2\pi $-periodicity condition for the function $\varphi _{l}(\theta )$
determines the energy levels $E_{l}$:

\begin{equation}
E_{l}=\frac{f(\pi )}{\pi }-2\mid h\mid \ M\ l.  \label{El}
\end{equation}
The normalization constant $C$ in (\ref{fii}) is chosen to yield 
\[
\langle \phi _{l}\mid \phi _{l^{\prime }}\rangle =\frac{\delta _{ll^{\prime
}}}{\Delta E}, 
\]
where $\Delta E=2\mid h\mid M$ is the interlevel distance.

\subsection{Decay rate}

The first-order correction to the false vacuum energy is trivial: $%
E^{(1)}=\langle 0_{+}\mid V\mid 0_{+}\rangle =\mathcal{N\,}\left| h\right|
\,M$, the second-order correction is given by 
\begin{equation}
E_{\text{ }}^{(2)}=-\Delta E\sum_{l}\frac{\mid \langle \phi _{l}\mid V\mid
0_{+}\rangle \mid ^{2}}{E_{l}}  \label{E2}
\end{equation}
To determine the decay rate of the false vacuum, the following trick is
used. We shift the excitation energy levels $E_{l}$ in (\ref{E2}) downwards
into the complex $E$-plane: $E_{l}\rightarrow $ $E_{l}-i\gamma $, where the
width $\gamma $ describes phenomenologically the decay rate of one-domain
states $\mid \phi _{l}\rangle $. Decay of these states should be caused by
the interaction term (\ref{vt}) in the same manner as the false vacuum decay
.

As the result, the metastable vacuum energy gains the imaginary part

\begin{equation}
\text{Im }E\simeq -\pi \ g(h)\mid \langle \phi _{l}\mid V\mid 0_{+}\rangle
\mid _{\text{ }E_{l}=0}^{2},  \label{ee}
\end{equation}
where 
\begin{equation}
g(h)=\text{Im}\cot \left[ \frac{f(\pi )-i\pi \gamma }{2\ \mid h\mid M}%
\right] .  \label{osci}
\end{equation}
The metastable vacuum relaxation rate $\Gamma $ is determined then in the
usual way (\ref{img}). It is evident from (\ref{ee}), (\ref{img}) that $%
\Gamma $ oscillates in $h^{-1}$ with the period $\Delta h^{-1}$ given by 
\begin{equation}
\Delta h^{-1}=2\pi M/f(\pi ).  \label{pe}
\end{equation}
These oscillations become considerable in the case of the resonant tunneling 
$\gamma \lesssim \Delta E$. In the opposite limit $\gamma \gg \Delta E$
oscillations in $h^{-1}$ vanish and relations (\ref{ee}), (\ref{img})
transform to the Fermi's golden rule:$^{(\citenum{LL,RS})}$ 
\begin{equation}
\Gamma =2\pi \mid \langle \phi _{l}\mid V\mid 0_{+}\rangle \mid _{\text{ }%
E_{l}=0}^{2}.  \label{GF}
\end{equation}

Let us now calculate the matrix element in (\ref{ee}): 
\begin{eqnarray}
\langle \phi _{l} &\mid &V\mid 0_{+}\rangle =\frac{1}{2}\int\limits_{-\pi
}^{\pi }\frac{d\theta }{2\pi }\ \phi _{l}^{*}(\theta )\ \langle 0_{+}\mid
\psi (-\theta )\,\psi (\theta )\ V\mid 0_{+}\rangle \simeq  \nonumber \\
\int\limits_{-\pi }^{\pi }\frac{d\theta }{2\pi }\ \varphi _{l}^{*}(\theta )\
\langle 0_{+} &\mid &\psi (-\theta )\,\psi (\theta )\ V\mid 0_{+}\rangle .
\label{mel}
\end{eqnarray}
Expanding operator $V$ in the $h$-power series 
\[
\langle 0_{+}\mid \psi (-\theta )\,\psi (\theta )\ V\mid 0_{+}\rangle
=\sum_{j=1}^{\infty }\langle 0_{+}\mid \psi (-\theta )\,\psi (\theta )\ 
\mathcal{H}^{(j)}\mid 0_{+}\rangle 
\]
and keeping in it only the leading ($j=1$) term one obtains from (\ref{v}), (%
\ref{hami}), (\ref{h1}): 
\begin{eqnarray*}
\langle 0_{+} &\mid &\psi (-\theta )\,\psi (\theta )\ V\mid 0_{+}\rangle
\simeq \langle 0_{+}\mid \psi (-\theta )\,\psi (\theta )\ \mathcal{H}%
^{(1)}\mid 0_{+}\rangle = \\
\langle 0_{+} &\mid &\psi (-\theta )\,\psi (\theta )\ U\mid 0_{+}\rangle \ 
\frac{\omega (\theta )}{\sinh \omega (\theta )}\ =-2\,i\!\!\ \mathcal{N}\mid
h\mid M\ \frac{d\ln \epsilon (\theta )}{d\theta }\cdot \frac{\omega (\theta )%
}{\sinh \omega (\theta )}\ .
\end{eqnarray*}
Thus, the matrix element (\ref{mel}) can be approximately represented as 
\begin{equation}
\langle \phi _{l}\mid V\mid 0_{+}\rangle \simeq -\,i\,\mathcal{N}\mid h\mid
M\int\limits_{-\pi }^{\pi }\frac{d\theta }{\pi }\frac{\omega (\theta )}{%
\sinh \omega (\theta )}\ \varphi _{l}^{*}(\theta )\ \frac{d\ln \epsilon
(\theta )}{d\theta }.  \label{mv}
\end{equation}

Substitution of (\ref{mv}), (\ref{ee}), (\ref{fii}) into (\ref{fr}) yields
the final expression for the imaginary part of the free energy $F_{ms}$ in
the limit $h\rightarrow -0$: 
\begin{equation}
\func{Im}F_{ms}\simeq \frac{\pi }{2}\mid H\mid M\ g(h)\left|
\int\limits_{-\pi }^{\pi }\frac{d\theta }{\pi }\frac{\omega (\theta )}{\sinh
\omega (\theta )}\ \frac{d\ln \epsilon (\theta )}{d\theta }\exp \left[ \frac{%
i\ f(\theta )}{2\mid h\mid M}\right] \ \right| ^{2}  \label{FIN}
\end{equation}
This expression generalizes formula (18) of reference$^{(\citenum{R})}$ to
arbitrary anisotropy $J_{1}/J_{2}$ and all temperatures $0<T<T_{c}$.

The last (exponent) factor of the integrand in (\ref{FIN}) oscillates with
high frequency in the considered case of small $\left| h\right| $.
Therefore, in the limit $\left| h\right| \rightarrow 0$, the integral in the
right-hand side of (\ref{FIN}) is determined by the saddle point of $%
f(\theta )$: $\theta =\theta _{1}\equiv -i\ln z_{1},\quad \omega (\theta
_{1})=0,$ and asymptotically equals to 
\begin{equation}
\func{Im}F_{ms}\simeq B\mid H\mid g(h)\exp \left[ -\frac{A}{\mid H\mid }%
\right] ,  \label{FIN2}
\end{equation}
where 
\begin{eqnarray}
A &=&\frac{\left| \ f(\theta _{1})\right| }{M\ \beta },\quad  \label{AA} \\
B &=&\frac{\pi \,M}{18}\ .  \label{B}
\end{eqnarray}

\section{Discussion}

First, let us establish equivalence of expressions (\ref{fd}) and (\ref{AA})
for the amplitude $A,$ which are given by the phenomenological droplet
theory and by our transfer-matrix calculations.

The droplet equilibrium shape in the $d=2$ Ising model is determined by the
equation 
\begin{equation}
a_{1}\cosh (\beta \lambda x_{1})+a_{2}\cosh (\beta \lambda x_{2})=1,
\label{sh}
\end{equation}
obtained by Zia and Avron.$^{(\citenum{ZA})}$ Here $x_{1},x_{2}$ denote
Descartes coordinates of a point on the droplet boundary, the scale
parameter $\lambda $ determines the droplet size, and 
\[
a_{1}=\frac{\tanh (2K_{2})}{\cosh (2K_{1})}\ ,\;\qquad a_{2}=\frac{\tanh
(2K_{1})}{\cosh (2K_{2})}. 
\]
It is remarkable, that equation (\ref{sh}) can be rewritten in terms of the
Ising model excitation spectrum (\ref{eps}) first obtained by Onsager:$^{(%
\citenum{Onsager})}$ 
\begin{equation}
x_{2}=\pm \frac{1}{\beta \lambda }\ \omega (i\,\beta \lambda \,x_{1}).
\label{!}
\end{equation}
Integrating in $x_{1}$ this equation we find the area of the
equilibrium-shaped droplet $S(\lambda )=W/\lambda ^{2}$, where 
\[
W=\frac{2}{\beta ^{2}}\left| \,f(\theta _{1})\right| . 
\]
It follows from Wulff's theorem$^{(\citenum{ZA})}$ that the surface energy $%
\Sigma (\lambda )$ also can be expressed in $W$: $\Sigma (\lambda
)=2W/\lambda $. Therefore, $\hat{\Sigma}^{2}=4W$, and 
\[
A=\frac{\beta \hat{\Sigma}^{2}}{8\ M}=\frac{\mid f(\theta _{1})\mid }{M\
\beta } 
\]
in exact agreement with (\ref{AA}).

Our expression $\pi \,M/18$ for the amplitude $B$ is the same as that
obtained previously in the extreme anisotropic limit.$^{(\citenum{R})}$ As
it was mentioned in the introduction, this expression is very close to the
Voloshin's result (\ref{Vol}). The latter is expected to be exact in the
critical region, where fluctuations are isotropic and universal. It is
likely, that the small discrepancy between (\ref{B}) and (\ref{Vol}) is
caused by approximations used in our modified perturbation theory. We hope
to clarify this question in future.

Let us compare obtained expressions for the amplitude $B$ with the numerical
results by Baker and Kim$^{(\citenum{1})}$ which they calculated for the
symmetric case $J_{1}=J_{2}\equiv J$ of the Ising model . These authors
considered the power series for the magnetization $M(h)$ at a fixed
temperature below $T_{c}$: 
\[
M(h)=M(0)-2\sum_{n=1}^{\infty }\left( -2h\right) ^{n}a_{n}, 
\]
and calculated numerically 12 coefficients $a_{n}$ in this expansion at $%
u=0.1\ u_{c}$, and 24 coefficients at $u=0.9\ u_{c}$. Here $u=\exp (-4\beta
J)$; $u_{c}=3-\sqrt{8}$ corresponds to the critical temperature. On the
other hand, Lowe and Wallace$^{(\citenum{LW})}$ demonstrated by use of the
dispersion relation, that equation (\ref{FF}) leads to the following
asymptotic formula for the coefficients $a_{n}$: 
\begin{equation}
a_{n}\stackunder{n\rightarrow \infty }{\rightarrow }\frac{B}{2\pi }\left(
2A\beta \right) ^{-n}\frac{(n+1)!}{n}.  \label{WL}
\end{equation}
So, the ratio 
\begin{equation}
R_{n\text{ }}=\frac{B\ (n+1)!}{2\pi \ n\ a_{n}\ (2A\beta )^{n}}  \label{rat}
\end{equation}
should approach to unity at large $n$, if we put in it the correct values of
amplitudes $A$ and $B$ . We plot in Fig. 2 this ratio, where coefficients $%
a_{n}$ were taken from paper$^{(\citenum{1})}$ by Baker and Kim. The left
pair of curves corresponds to the low temperature case $u=0.1\ u_{c},$ the
right pair of curves corresponds to the higher temperature $u=0.9\ u_{c}$.
The amplitude $A$ in (\ref{rat}) is taken from (\ref{AA}). Solid and dashed
curves differ by choice of the amplitude $B$ in (\ref{rat}). In solid
curves, it is chosen as $B=\pi M/18$ according to our result (\ref{B}); in
dashed curves $B=M/(2\pi )$ according to Voloshin's result.$^{(%
\citenum{Voloshin2})}$ All four curves in Fig. 2 seems to stabilize at large 
$n$ to the values, which are rather close to unity. This indicates a
remarkable good agreement of numerical results$^{(\citenum{1})}$ with
expressions (\ref{Vol}) or (\ref{AA}). Though, agreement with Voloshin's
value seems somewhat better, saturation in curves is not achieved, and
further numerical calculation are desirable to distinguish between (\ref{Vol}%
) and (\ref{AA}).

Expression (\ref{FIN2}) differs from (\ref{FF}) by the oscillating factor $%
g(h)$. We interpret this factor as the correction coursed by the
discrete-lattice effects. These oscillation being negligible in the critical
region, may be significant at low temperatures, especially in the presence
of strong anisotropy.$^{(\citenum{O})}$ The period of oscillations in $%
h^{-1} $ is given by (\ref{pe}). It is plotted in Fig. 3 in the symmetric
case $K_{1}=K_{2}\equiv K$. Such oscillations with period $\Delta
h^{-1}\approx 1/2 $ were observed in numerical constrained transfer matrix
calculations by G\"{u}nther \textit{et al.}$^{(\citenum{GRN})}$\textit{\ }at 
$K=1$. This period agrees well with our value $\Delta h^{-1}=0.494891,$
which follows from (\ref{pe})$.$

\section{Acknowledgments}

I would like to thank Professor Royce Zia for helpful correspondence.

This work is supported by the Fund of Fundamental Investigations of Republic
of Belarus.

\appendix

\section{Fermionization\label{apa}}

In this Appendix we present fermionic representations of spin operators,
which are used in Section \ref{tr}. Consideration is restricted to the
ferromagnetic phase $T<T_{c}$ in the thermodynamic limit $\mathcal{N}%
\rightarrow \infty $. In this limit the Jordan-Wigner transformation can be
written as$^{(\citenum{JM})}$ 
\begin{equation}
P_{n}=\sigma _{n}^{3}\ \sigma _{n-1}^{1}\ \sigma _{n-2}^{1}\ \ldots ,\quad
Q_{n}=-i\,\sigma _{n}^{2}\ \sigma _{n-1}^{1}\ \sigma _{n-2}^{1}\ \ldots .
\label{JW}
\end{equation}
Here $P_{n}$ and $Q_{n}$ are the fermionic operators satisfying the
following anticommutational relations: 
\[
\left\{ P_{n}\ ,P_{n^{\prime }}\right\} =2\delta _{nn^{\prime }},\qquad
\left\{ Q_{n}\ ,Q_{n^{\prime }}\right\} =-2\delta _{nn^{\prime }},\qquad
\left\{ P_{n}\ ,Q_{n^{\prime }}\right\} =0. 
\]
Let us define the another set of fermionic operators $p_{n}$, $q_{n}$, which
are related with $P_{n}$ , $Q_{n}$ by the duality transformation:$^{(%
\citenum{Kog})}$ 
\begin{equation}
p_{n}=i\,Q_{n},\quad q_{n}=-i\,P_{n+1}.  \label{du}
\end{equation}
Operators $p_{n}$, $q_{n}$ obey the same anticommutational relations as $%
P_{n}$ , $Q_{n},$ span an orthogonal space of free fermion field, and
generate the Clifford algebra$^{(\citenum{JM})}.$

Fermionic creation and annihilation operators $\psi (\theta ),\psi ^{\dagger
}(\theta )$ introduced in Section \ref{tr} are related with $p_{n}$, $q_{n}$
by 
\begin{eqnarray}
2i\psi (\theta ) &=&e^{-i\alpha (\theta )}\ p(\theta )-\ e^{i\alpha (\theta
)}q(\theta ),  \label{psi} \\
2i\psi ^{\dagger }(-\theta ) &=&e^{-i\alpha (\theta )}p(\theta )+e^{i\alpha
(\theta )}q(\theta ),  \nonumber \\
p(\theta ) &=&\sum_{n\in \Bbb{Z}}e^{-in\theta }\ p_{n},\quad q(\theta
)=\sum_{n\in \Bbb{Z}}e^{-in\theta }\ q_{n},  \nonumber
\end{eqnarray}
where 
\begin{eqnarray}
e^{2i\alpha (\theta )} &=&-\tanh K_{2}\ \left[ \frac{(e^{i\theta
}-z_{1})(e^{i\theta }-z_{2}^{-1})}{(e^{i\theta }-z_{2})(e^{i\theta
}-z_{1}^{-1})}\right] ^{1/2},  \label{alfa} \\
e^{2i\alpha (0)} &=&-1,  \nonumber
\end{eqnarray}
parameters $z_{1},\ z_{2}$ are defined by (\ref{zz}).

Relations (\ref{JW}), (\ref{du}) express operators $p_{n}$, $q_{n}$ in terms
of Pauli matrices. The inverse transformation reads as 
\begin{eqnarray}
\ \sigma _{n}^{1} &=&p_{n}\ q_{n-1},\quad  \nonumber \\
\sigma _{n}^{2} &=&p_{n}\ \left( p_{n-1\ }q_{n-2}\right) \left( p_{n-2}\
q_{n-3}\right) \ldots ,  \label{sig2} \\
\sigma _{n}^{3} &=&i\,q_{n-1}\ \left( p_{n-1}\ q_{n-2}\right) \left(
p_{n-2}\ q_{n-3}\right) \ldots .  \label{sig3}
\end{eqnarray}
Brackets in equations (\ref{sig2}), (\ref{sig3}) are shown to indicate, that 
$\sigma _{n}^{2}$ and $\sigma _{n}^{3}$ are the products of odd number of
fermionic operators, i.e. $\sigma _{n}^{2}$ and $\sigma _{n}^{3}$ are the
odd elements of the Clifford group.

Following Jimbo \textit{et al.}$^{(\citenum{JM})}$\textit{, }let us
introduce operator $\bar{\sigma}_{n}^{3},$ which represent $\sigma _{n}^{3}$
under the boundary condition $\sigma _{n}^{3}\rightarrow 1$ for $%
n\rightarrow -\infty $: 
\[
\bar{\sigma}_{n}^{3}=\,\left( q_{n-1}\ p_{n-1}\right) \left(
q_{n-2}p_{n-2}\right) \ldots . 
\]
This operator is an even element of the Clifford group. Due to the obvious
identity $\bar{\sigma}_{n}^{3}\bar{\sigma}_{n^{\prime }}^{3}=\sigma
_{n}^{3}\sigma _{n^{\prime }}^{3}$ , operators $\sigma _{n}^{3}$ and $\ \bar{%
\sigma}_{n}^{3}$ produce the same correlation functions, which makes
reasonable to identify them in the thermodynamic limit. Accordingly, we
shall replace $\sigma _{n}^{3}$ by$\ \bar{\sigma}_{n}^{3}$ in the operator $%
U $ : 
\begin{equation}
U\mapsto h\sum_{n=1}^{\mathcal{N}}\,\bar{\sigma}_{n}^{3}.  \label{rep}
\end{equation}

Operators $\bar{\sigma}_{n}^{3}$ are characterized up to $\pm 1$ factor by
the following commutation relations: 
\begin{mathletters}
\begin{equation}
\bar{\sigma}_{n}^{3}\ q_{n^{\prime }}\ \bar{\sigma}_{n}^{3}=\varkappa
(n-n^{\prime })\ q_{n^{\prime }},\quad \bar{\sigma}_{n}^{3}\ p_{n^{\prime
}}\ \bar{\sigma}_{n}^{3}=\varkappa (n-n^{\prime })\ p_{n^{\prime }},
\label{ort}
\end{equation}
where 
\end{mathletters}
\[
\varkappa (n)=\left\{ 
\begin{array}{ll}
1 & \text{for }n\geq 0 \\ 
-1 & \text{for }n<0
\end{array}
\right. . 
\]
Thus, $\bar{\sigma}_{n}^{3}$ induces a linear orthogonal transformation of
the linear space of free fermions. As it was shown by Jimbo \textit{et al.}$%
^{(\citenum{JM})}$\textit{\ }in Appendix 1, such an operator can be
expressed as the normally ordered exponent 
\begin{equation}
\bar{\sigma}_{n}^{3}=\left\langle \bar{\sigma}_{n}^{3}\right\rangle :\exp
\left( \rho _{n}/2\right) :.  \label{normal}
\end{equation}
Here operator $\rho _{n}$ is quadratic in free fermionic variables $p_{n}$
and $q_{n}$, $\left\langle \bar{\sigma}_{n}^{3}\right\rangle $ is the vacuum
expectation value of $\bar{\sigma}_{n}^{3}$, i.e. the spontaneous
magnetization $\left\langle \bar{\sigma}_{n}^{3}\right\rangle =M=\left[
1-k^{2}\right] ^{1/8}$. The explicit expression for $\rho _{n}$ reads as 
\begin{eqnarray}
\frac{\rho _{n}}{2} &=&-\sum_{j<\,n}\psi _{j}^{(+)}\ \psi _{j}^{(-)},
\label{rr} \\
\psi _{j}^{(+)} &=&-\int_{-\pi }^{\pi }\frac{d\theta }{2\pi }\ e^{ij\theta
}\ U_{+}(-\theta )\ p(\theta )\ ,  \nonumber \\
\psi _{j}^{(-)} &=&\int_{-\pi }^{\pi }\frac{d\theta }{2\pi }\ e^{ij\theta }\
U_{-}(\theta )\ q(\theta ),  \nonumber
\end{eqnarray}
where the functions 
\[
U_{\pm }(\theta )=\left[ \epsilon (\theta )\right] ^{\mp 1}\ \exp \left[
i\alpha (\theta )\right] 
\]
provide the Wiener-Hopf factorization of $\exp \left[ 2i\alpha (\theta
)\right] $: 
\[
\exp \left[ 2i\alpha (\theta )\right] =U_{+}(\theta )\ U_{-}(\theta ). 
\]
Functions $U_{+}(\theta )$ and $U_{\_}(\theta )$ are analytical in $%
z=e^{i\theta }$ outside and inside the unit circle, respectively. Rewriting $%
\psi _{j}^{(+)}$ and $\psi _{j}^{(-)}$ in terms of creation and annihilation
operators $\psi ^{\dagger }(\theta ),\,\psi (\theta )$ by use of (\ref{psi})
one obtains from (\ref{rep}), (\ref{normal}), (\ref{rr}) the desired
fermionic representation (\ref{v}) of the $U$ operator.

In deriving (\ref{rr}) we have chosen the free fermion basis $p(\theta ),\
q(\theta )$ and applied the theorem, presented by Jimbo \textit{et al. }in
pages 137, 138 of their article$^{(\citenum{JM})}$. In our case, the kernel
functions for matrices $P,$ $E$ introduced in this theorem read as (compare
with equations (3.15), (3.16) in the same article): 
\begin{eqnarray*}
P(\theta ,\theta ^{\prime }) &=&\frac{e^{i(n-1)(\theta -\theta ^{\prime })}}{%
1-e^{-i(\theta -\theta ^{\prime }-i\,0)}}, \\
E(\theta ,\theta ^{\prime }) &=&\left( 
\begin{array}{cc}
0 & \exp \left[ -2i\alpha (\theta )\right] \\ 
\exp \left[ 2i\alpha (\theta )\right] & 0
\end{array}
\right) 2\pi \delta (\theta -\theta ^{\prime }).
\end{eqnarray*}

To illustrate convenience of representations (\ref{v}), (\ref{normal}), (\ref
{rr}), we shall apply them to derive a compact Fredholm determinant formula
for the zero-field correlation function $\left\langle \sigma _{0,0}\ \sigma
_{m,n}\right\rangle $ in the ferromagnetic phase. Let us first write this
correlation function by use of (\ref{normal}), (\ref{v}) in the form 
\begin{eqnarray}
\left\langle \sigma _{0,0}\ \sigma _{m,n}\right\rangle  &=&\langle 0_{+}\mid 
\bar{\sigma}_{0}^{3}\exp (-m\mathcal{H}^{(0)})\ \bar{\sigma}_{n}^{3}\mid
0_{+}\rangle =  \nonumber \\
M^{2}\ \langle 0_{+} &\mid &:\exp \left( \rho _{0}/2\right) :\exp \left( -%
\mathcal{H}^{(0)}m\right) :\exp \left( \rho _{n}/2\right) :\mid 0_{+}\rangle
,  \label{me}
\end{eqnarray}
where 
\begin{eqnarray*}
\frac{\rho _{0}}{2} &\mapsto &\frac{1}{2}\iint_{-\pi }^{\pi }\frac{d\theta
_{1}d\theta _{2}}{\left( 2\pi \right) ^{2}}\ e^{-\frac{i}{2}(\pi +\theta
_{1}+\theta _{2})}\ D(\theta _{1},\theta _{2})\ \psi (\theta _{1})\psi
(\theta _{2}), \\
\frac{\rho _{n}}{2} &\mapsto &\frac{1}{2}\iint_{-\pi }^{\pi }\frac{d\theta
_{1}d\theta _{2}}{\left( 2\pi \right) ^{2}}\ e^{\frac{i}{2}(\pi +\theta
_{1}+\theta _{2})}\ e^{-in(\theta _{1}+\theta _{2})}\ D(\theta _{1},\theta
_{2})\ \psi ^{\dagger }(\theta _{1})\ \psi ^{\dagger }(\theta _{2}), \\
D(\theta _{1},\theta _{2}) &=&\frac{1}{2\sin \left[ (\theta _{1}+\theta
_{2})/2\right] }\left[ \frac{\epsilon (\theta _{1})}{\epsilon (\theta _{2})}-%
\frac{\epsilon (\theta _{2})}{\epsilon (\theta _{1})}\right] .
\end{eqnarray*}
We have dropped all creation operators $\psi ^{\dagger }$ in $\rho _{0}/2$,
and all annihilation operators $\psi $ in $\rho _{n}/2$, since the normally
ordered exponents of $\rho $ and $\rho _{0\text{ }}$act in (\ref{me}) on the
vacuum states. In the well-known holomorphic representation$^{(\citenum{SF})}
$ of fermionic operators, the matrix element (\ref{me}) takes the form of
the Gaussian continual integral over Grassnann variables, which integration
yields immediately: 
\begin{eqnarray}
\left\langle \sigma _{0,0}\ \sigma _{m,n}\right\rangle  &=&M^{2}\det \left(
1-D_{mn}\right) =  \label{det1} \\
&&M^{2}\exp \left( -\sum_{j=1}^{\infty }\frac{\mathrm{Sp}D_{mn}^{2j}}{2j}%
\right)   \label{2dd}
\end{eqnarray}
Here $D_{mn}$ denotes a linear integral operator acting on the function $%
f(\theta )$ as follows: 
\[
D_{mn}\ f=\int\limits_{-\pi }^{\pi }\frac{d\theta ^{\prime }}{2\pi }\
D(\theta ,\theta ^{\prime })\exp \left\{ -\frac{in(\theta +\theta ^{\prime })%
}{2}-\frac{m\left[ \omega (\theta )+\omega (\theta ^{\prime })\right] }{2}%
\right\} \ f(\theta ^{\prime }).
\]
Equation (\ref{det1}) is a compact forms of the well-known exact
representation of the two-point correlation function in the Ising model
obtained by Wu \textit{et al.}$^{(\citenum{Wu})}$ (see equations
(2.9)-(2.13) in the referred article). The latter representation can be
reduced to (\ref{2dd}) by explicit integration in the right-hand side of
equation (2.12) in $\phi _{1},\phi _{3},\phi _{5,}\ldots $.

\begin{center}
{\LARGE Figure Captions}
\end{center}

\begin{quote}
\textbf{Figure 1: }Free energy continuation paths from the positive real
axis $H>0$ to the cut going along the negative real axis $H<0$.

\textbf{Figure 2: }Plot of $R_{n}$ given by (\ref{rat}) versus $n$.
Coefficients $a_{n}$ in (\ref{rat}) are taken from reference$^{(\citenum{1}%
)} $; amplitude $A$ is taken from (\ref{AA}); amplitude $B$ is taken either
from (\ref{B}) (solid curves), or from (\ref{Vol}) (dashed curves). Two left
curves correspond to $u=0.1u_{c}$, two right curves correspond to $%
u=0.9u_{c} $.\newline

\textbf{Figure 3: }Oscillation period $\Delta h^{-1}$ versus $K$ in the
symmetric case.
\end{quote}

\end{document}